\documentclass[%
aip,
amsmath,amssymb,
reprint,%
]{revtex4-1}
\usepackage{amssymb}
\usepackage{amsfonts}
\usepackage{amsmath}
\usepackage{graphicx}%
\begin{document}
\title{Is it possible to suspend the spread of an epidemic infection? The dynamic Monte Carlo approach.}
\author{Gennadiy Burlak}
\affiliation{Centro de Investigaci\'on en Ingenier\'ia y Ciencias Aplicadas, Universidad Aut\'onoma del Estado de Morelos. 
	Av. Universidad 1001, Col. Chamilpa, C.P. 62210, Cuernavaca, Morelos, M\'exico}

\begin{abstract}
We study a dynamics of the epidemiological infection spreading at different values of the risk factor 
$\beta$ (a control parameter) with the using of dynamic Monte Carlo approach (DMC). In our toy model, the infection transmits 
due to contacts of randomly moving individuals. We show that the behavior of recovereds  critically depends on 
the $\beta$ value. For sub-critical values $\beta<\beta_{c}\sim 0.6$, the number of infected cases asymptotically 
converges to zero, such that for a moderate  risk factor the infection may disappear with time. Our simulations 
shown that over time, the properties of such a system asymptotically become close to the critical transition in 2D 
percolation system. We also analyzed an extended system, which includes two additional parameters: the limits of taking on/off 
quarantine state. It is found that the early quarantine off does result in the irregular (with positive Lyapunov exponent) 
oscillatory dynamics  of infection. If the lower limit of the quarantine off is small enough, the recovery dynamics acquirers 
a characteristic nonmonotonic shape with several damped peaks. The dynamics of infection spreading in case of the individuals 
with immunity is studied too.
\end{abstract}
\maketitle

\textbf{The dangerous trends of the coronavirus spreading throughout the world gives rise to numerous investigations in wide scientific spectrum. We study the spread of epidemiological infection at different values ​​of the risk factors $beta$ with the use of the dynamic Monte Carlo (DMC) method. In such a model, it is accepted that the infection is transmitted through the contacts of randomly moving individuals. We show that the quantity of recovered critically dependent on the value $\beta$. It is remarkably that for sub-critical values $\beta <\beta_{c} \sim 0.6$ the number of infected cases asymptotically converges to zero, so that for a moderate risk factor, the infection can quickly disappear. Our calculations showed that such a dynamic property of the system (asymptotically) is associated with the critical behavior in 2D percolation medium. We also analyze an extended system, which is currently widely used to prevent the spread of the virus and includes quarantine on/off settings. It was revealed that early exit from the quarantine state leads to irregular oscillatory dynamics of infection. However, when the lower limit of quarantine off is small enough, the dynamics of infection acquires a characteristic non-monotonic shape with several damped peaks. Comparison of quarantine and immunity factors shows that in the case of immunity, the complete recovery occurs faster than in a quarantine mode.}

\section{Introduction}
The dangerous dynamics of the coronavirus spread throughout the world gives rise to numerous studies in a wide scientific spectrum. Improving known epidemic models and developing new models are complicated tasks because the lack of verified statistics on the infection spread and disease dynamics. Unstable predictability of infection, ambiguity with drugs \cite{Corey:2020}, uncertainty regarding the immunity of disease \cite{Nature:2020}, and other factors (such as a viral mutation) make it difficult to predict the dynamics of pandemic. 
This may relay to some mathematical models, which depend on a significant number of free (statistically-driven) parameters. The known models of the SIR family give solutions in a form of smooth functions\cite{Choisy:2007} (solutions of differential equations) that only indirectly include important random factors. 
Naturally, that in such a situation, most statistically reliable forecasts are obtained by methods based on the direct application of the central limit theorem with a predicted error of $1/\sqrt N$, see \cite{Barmparis:2020a},\cite{Eugene:2020a},\cite{Fandaou:2020a} and references therein. 
In this paper, we propose the use of the dynamic Monte Carlo (DMC) method that self-consistently includes various dynamic random factors. Such a technique was previously used to study the processes associated with aggregation, viscous flow properties, the formation of biological structures, and allows to scale the associated geometric and dynamic quantities that characterize these phenomena \cite{Vicsek:1995},\cite{Solon:2015},\cite{Solon:2015a}. 
In our study, as a control (free) parameter, we use the generalized risk factor $\beta$, which includes some of the factors mentioned above in an integral form. In our 2D toy model, the transmission of infection occurs due to contacts of randomly moving individuals, that determines the complex dynamics of the infection spread and various critical aspects of such a dynamics.
The paper is organized as follows. In Section 2, we formulate our approach and examine the behavior of infected individuals (order parameter $A(t,\beta)$), which, as it turns out latter, critically depends on the value $\beta$. It also is discussed the similarity of the asymptotic behavior of the infection dynamics with the critical phase transition in a two-dimensional (2D) percolation system. In the next Section, we analyze the dynamic properties of the extended system, where we deal with two additional parameters which allow to on/off the quarantine state. The next Section, contains the study of dynamics of the infection spreading and the formation of immunity for infected individuals. The last Section summarizes our conclusions.

\section{Dynamic Monte Carlo simulations}
First we explain which the properties of dynamic Monte Carlo (DMC) approach we deal with. In order to study the infection dynamic in epidemiological system (that is far from equilibrium) the DMC method is used, that allows investigation both temporal and spatial properties by the numerical simulations.
As a toy model, we choose a 2D $L\times L$ (where $L$ is
size) bounded system that contains a disordered population of $N$ individuals.
Following the classifications commonly known from  SIR model \cite{Choisy:2007} in our DMC model  we divide the host population into a set of distinct categories, 
according to its epidemiological status, that are susceptible (S), currently infectious (I ), and recovered (R). The total size of the host population is then $ N = S + I + R $ and all the individuals are born in the susceptible 
category\cite{Choisy:2007}. Following the actual situations we assume that initially the  maternally derived immunity is clear. (The effect of immunity is studied in the following sections). Upon contact with infectious individuals, the susceptibles may get infected and move into the infectious category. To apply the DMC approach it is constructed the Person class (individual, alias object) that encapsulates properties of a randomly placed and moving individual and contains the following significant attributes%
\begin{equation}
\{x,y,v_{x},v_{y},I,M\},\label{attrib}%
\end{equation}
where $x,y$ are the components of position, $v_{x},v_{y}$ are the components
of velocity, the parameters $I,M$ describe the states: infected/uninfected
and immunized/non-immunized, respectively. The list of Persons that represents a total host population is used in our DMC simulations.

 One of the underlying reasons why epidemiological systems exhibit variation is due to a
 different way that the individuals in a population have contact with each other.  
 In our DMC simulation we assume that the spreading (transmission) of the infection occurs 
 because of random contacts for moving individuals.   
 
To do that in DMC simulations we use the following strategy. (i) Any contact can occur only between two nearest individuals. (ii) At any contact, the state of an infected transmits to the other contact person. But the infected one can still be recovered with probability $1-\beta$ (recall that $\beta \in [0,1]$ is a risk factor). This means that if $\beta \lesssim 1$, the probability to a recovery is small. 

\begin{figure}[ptb]%
 	\centering
 	\includegraphics[
 	width=0.45\textwidth
 	]%
 	{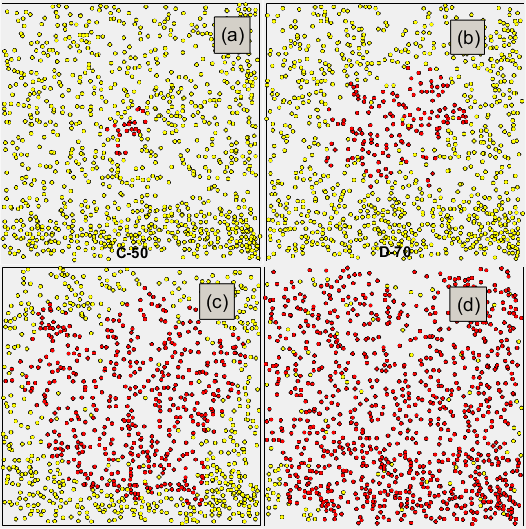}%
 	\caption{(Color on line.) The snapshots ($N=1000, \beta=0.95$, $L=400$) show the dynamics of the infection spreading at: (a) t=10, (b) t=30, (c) t=50, (d) t=70. We observe that for shown case at $t=70$ nearly all the individuals are infected.	
 	}%
 	\label{Pic_Fig1a}%
 \end{figure}

To take an advantage of the visualizations at applying the DMC technique, we allow each object to have a visual representation, which is a yellow circle (non-immunized individual), a green circle (immunized but not infected individual) and a red circle (infected individual), see Fig.\ref{Pic_Fig1a}.  
We used the interaction radius $r$ as the unit scale to measure distances (used $r = 6$, see Fig.\ref{Pic_Fig1a}) between the individuals, while the time unit $\Delta t = 1$ was the 
time interval between two updatings of the directions and positions. 
In our simulations we used the
simplest initial conditions: at time $t=0$ the positions and velocities for all the $N$ individuals are randomly distributed.
We use the velocity scale such that random $v_x, v_y \in [2,10]$ for which the
individuals always interact with their actual neighbors and move fast enough to change the configuration after a few
updates. According to our simulations, the variation of
actual interval of values of $v_x, v_y$ does not affect the results. 
We also investigated the cases when the basic parameter of the model, the density
$\rho = N/L^2$ is slightly varied. 

   
   When the simulation time runs a lot contacts occur between nearest randomly moving persons that leads to fast and uncontrollable transfer of infection between many individuals, see Fig. \ref{Pic_Fig1a}.  
\begin{figure}[ptb]%
	\centering
	\includegraphics[
	width=0.5\textwidth
	]%
	{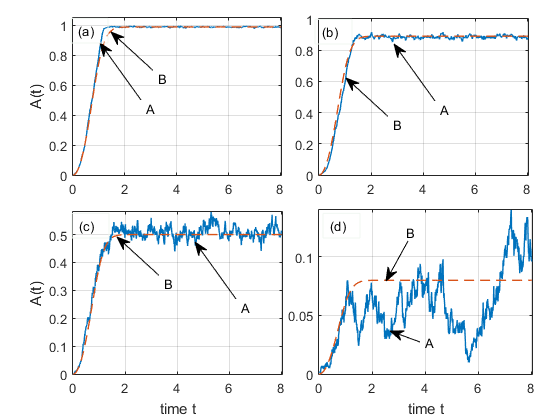}%
	\caption{(Color on line.) The dynamics of the infections spreading
		coefficient (order parameter) $A=I/N$ for times $t<8$ at different values of the risk factor (control parameter) $\beta$. [In this figure the abscissa axis shows the fitting time $0.01t$]
		The blue
		lines (arrows A) show the numerical simulations (DMC) data, while red lines
		(arrows B) display the fitting function Eq.(\ref{fitting}), where only $a_{0}$ coefficient changes considerably at $\beta$\ variation: (a) shows
		case $\beta=0.99$, (b) $\beta=0.90$, (c) $\beta=0.80$, (d) $\beta = 0.60$. At
		$\beta<0.60$ the DMC solution rapidly converges to 0 ($A=0$). This means that for
		$\beta<\beta_{c}\approx0.60$ all the infected individuals will be recovered up to $t=8$.
	}%
	\label{Pic_Fig2a}%
\end{figure}
It is of great interest to investigate the temporal infection dynamics at
various risk factors $\beta$. Such a dynamics of the infections
spreading (coefficient $A(t)=I(t)/N$) as function of time $t$ is displayed in Fig.\ref{Pic_Fig2a}.
Since $A(t)$ is a random-valued function we will fit (see \cite{Press:2002}) $A(t)$ by a suitable
fitting function that is chosen as 
\begin{equation}
f(t)=a_{0}t^{a_{1}}\tanh(t^{a_{2}}),\label{fitting}%
\end{equation}
where $a_{0,1,2}$ are the fitting coefficients.  We found that $a_{1}$ is very small $\sim 10^{-5}$
and $a_{2}\simeq2$ for all the cases, but the amplitude $a_{0}$ changes considerably at
the $\beta$\ variation. In Fig.\ref{Pic_Fig2a} the blue lines
show the numerical simulations (DMC) data, while the red lines display the
fitting function $f(t)$. Fig.\ref{Pic_Fig2a} (a) shows the case $\beta=0.99$, (b) $\beta=0.90$,
(c) $\beta=0.80$, (d) $\beta=0.60$. We indicate a \textit{remarkable} observation that for $\beta<0.60$ the system asymptotically converges to a
trivial solution with $A\simeq a_0=0$ already for $t\simeq 6$.%
\begin{figure}[ptb]%
	\centering
	\includegraphics[
	width=0.5\textwidth
	]%
	{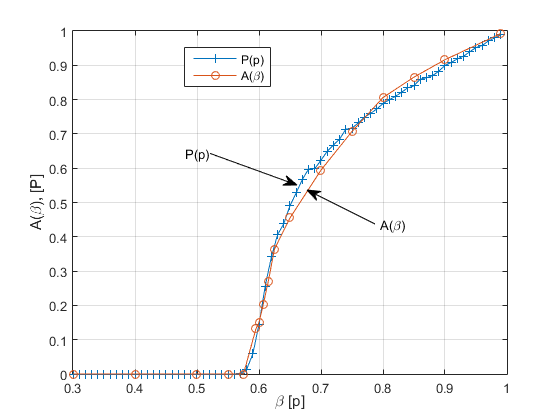}%
	\caption{(Color on line.) The comparison of the order parameter functions for the infection spreading $A(\beta)$ and the order parameter for 2D percolating $P(p)$, where $p$ is is the occupation probability of defect state\cite{Grimmett:1999a}. Red line shows the dependence of (slightly tailored) amplitude parameter the fitting function $a_0$ corresponding 
		to $A=I/N$ at fixed time $t=8$ as function
		of the risk factor $\beta$. Blue line shows the dependence of the order		
		parameters $P(p)$ for 2D percolating material as function of the occupation probability $p$. We observe that both curves are very close and  the phase transitions to
		infected/percolating state occurs similarly at $\beta_{c}\simeq p_{c}\simeq0.6$ for both cases.}%
	\label{Pic_Fig3a}%
\end{figure}
Such observation leads to an interesting assumption that the studied dynamics of the infection spread can (asymptotically) be associated with a critical transition in the two-dimensional (2D) percolation system, that occurs when the occupation probability of defects is  $p_{c}=0.594$ \cite{Isichenko:1992a},\cite{Grimmett:1999a},\cite{Stauffer:1992},\cite{Burlak:2009a},\cite{Burlak:2015} see Fig.\ref{Pic_Fig3a}. Such  an assumption is studied in the following Section.

\section{Critical value of the risk factor}
Fig. \ref{Pic_Fig3a} displays a comparison of above mentioned dependencies. In Fig.
\ref{Pic_Fig3a} the red line shows the dependence $a_0(\beta)$ (see Fig.\ref{Pic_Fig2a}) associated with the infecting parameter $A(\beta)=I/N$, and the blue
line shows the percolating order parameters $P(p)$ as function of the
occupation defect probability $p$. We observe that both dependencies are in excellent agreement and at $\beta_{c}\simeq p_{c}\simeq0.6$ the phase transition to infected/percolating state occurs. From Fig. \ref{Pic_Fig3a} we can assume that the parameter
$A(\beta)$ can be mentioned further as an order parameter (similarly $P(p)$).
This results that the formalism of the percolation critical  percolating phase transition \cite{Isichenko:1992a},\cite{Grimmett:1999a},\cite{Stauffer:1992} 
can be  applied to investigation the asymptotic of infection spreading at various $\beta$. 
On the other hand, good agreement between the results of DMC modeling and the critical transition in 2D percolation system shows the general applicability of DMC approach to analyze the dynamics of infection spread in the epidemiological system.

\section{The extension of model}
\subsection{Quarantine regime}
Mass infection shown in Fig. \ref{Pic_Fig1a} is an extremely dangerous and highly undesirable scenario for the development of epidemiological situation. This Section discusses the extension of the model, which in principle allows
suspending this trend. One of the simple solutions proposed recently is introducing the quarantine by localizing of infected individuals in order to significantly reduce the number of contacts that leads to the transmission of infection. This can be modeled  by setting $v_x=v_y=0$ 
for infected individuals and ignoring all the contacts with them in our approach.
We call such a regime of simulation as a quarantine mode. In order to do this we 
introduce two new parameters into the model, $A_{\max}$ (the infection level 
when the quarantine is automatically turned on), and $A_{\min}$ (the infection
level when the quarantine is turned off).   
   
\begin{figure}[ptb]%
\centering
\includegraphics[
width=0.5\textwidth
]%
{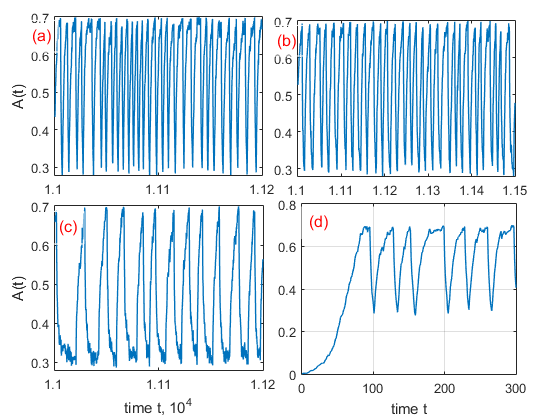}%
\caption{(Color on line.) The dynamics of infection parameter $A$ in quarantine mode with $A_{\max}=0.7$, $A_{\min}=0.36$ at moderate values $\beta$, (a) $\beta=0.78$, ( b) $\beta=0.80$, and (c) $\beta=0.82$; panel (d) shows the typical dynamics $A$ at initial times. We observe the generation of irregular oscillations of $A$ with large amplitudes between $A_{\max}$ and $A_{\min}$. We calculated (by the method \cite{Alanwolf:1985a}) that the Lyapunov exponent for such irregular oscillations is about $0.3$.
}%
\label{Pic_Fig4a}%
\end{figure}
   
Fig. \ref{Pic_Fig4a} shows the dynamics of infections $A(t)$ in quarantine mode with $A_{\max}=0.7$, $A_{\min}=0.36$ at moderate values $\beta$, (a) $\beta=0.78$, (b) $\beta=0.80$, and (c) $\beta=0.82$; panel (d) shows the typical dynamics $A$ at initial times. For such parameters from Fig.\ref{Pic_Fig4a} we observe that the system transmits to unexpected dynamic state: the generation of irregular oscillations of $A$ with large amplitudes between $A_{\max}$ and $A_{\min}$. We have calculated (by the method \cite{Alanwolf:1985a}) that the Lyapunov exponent for such irregular oscillations is about $0.3$. 
This means that if quarantine is turned off too early, the growth of infections suppressed, but the system goes into dynamic mode with irregular oscillations. In this case, a significant number of infected and recovered individuals can be re-infected, therefore, a full recovery does not occur.
   
\begin{figure}[ptb]%
	\centering
	\includegraphics[
	width=0.5\textwidth
	]%
	{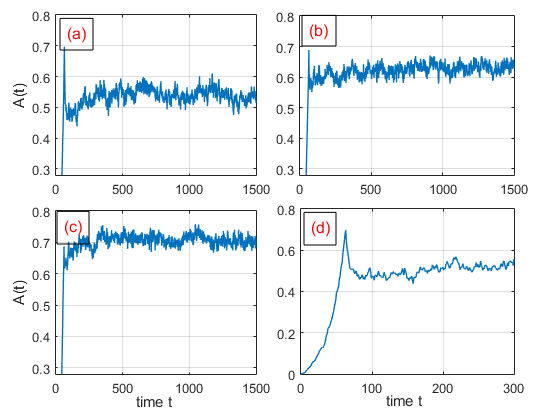}%
	\caption{(Color on line.) The same quarantine case as in Fig.\ref{Pic_Fig4a} but for large the risk factor $\beta$: (a)  $\beta=0.88$, (b)  $\beta=0.90$, (c)  $\beta=0.92$, (d)  $\beta=0.88$ for small times. We observe that for large $\beta$ the evolution of the infections has monotonic shape (with small random variations) but without the large oscillations shown in Fig.\ref{Pic_Fig4a}. One can see that the dynamics $A(t)$ in (b) for $\beta=0.90$ is already suppressed and considerably differs with respect to situation without the quarantine shown in Fig.\ref{Pic_Fig2a} (b).
	}%
	\label{Pic_Fig5a}%
\end{figure}
   
Fig.\ref{Pic_Fig5a} shows the infections dynamics $A(t)$ in the quarantine mode but for large the risk factors $\beta$: (a)  $\beta=0.88$, (b)  $\beta=0.90$, (c)  $\beta=0.92$, (d)  $\beta=0.88$. We observe that for large $\beta$ the evolution of the infections has monotonic shape (with small random variations) but without the oscillations as in Fig.\ref{Pic_Fig4a}. However the dynamics $A(t)$ in (b) for $\beta=0.90$ is already suppressed and strongly differs with respect to a case without the quarantine shown in  Fig.\ref{Pic_Fig2a} (b).

\subsection{The immunity}
\begin{figure}[ptb]%
\centering
\includegraphics[
width=0.5\textwidth
]%
{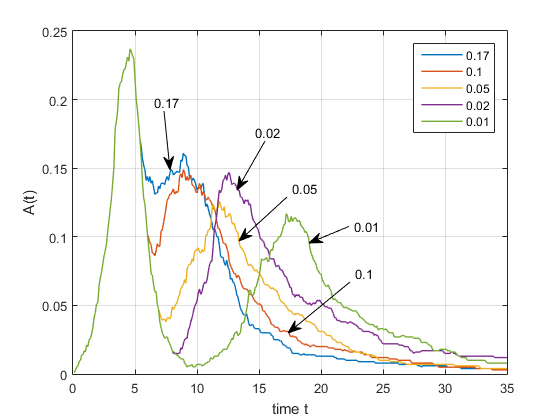}%
\caption{(Color on line.) The dynamics of recovery in presence of immunity
	in the quarantine mode for the parameters $\beta = 0.94 $, $ A_{\max} = 0.24 $ for different $A {\min} =  0.17, 0.1, 0.05, 0.02, 0.01$. One can see that after the high peak, the oscillations rapidly decay that leads to a decrease of infections (the order parameter $A(t) $ rapidly decreases).}%
\label{Pic_Fig6a}%
\end{figure}

Although actually there are  no reliable statistics for the congenital or acquired immunity for persons (for animals see Ref.\cite{Chandrashekar:2020}), in this Section we analyze this aspect in framework of our model. Following the Ref.\cite{Choisy:2007}, we assume that in the host population there is no innate immunity to the virus. But we suppose that the persons (at least a large majority) will acquire this immunity, as is usually the case. To this end, in our model we use the parameter $M$, see Eq. (\ref{attrib}). Following \cite{Choisy:2007}, we assume that this parameter acquires a non-zero value (the presence of immunity) only after first infection and recovery. Re-infection no longer occurs even at contacts with infected persons.
Fig. \ref{Pic_Fig6a} shows the dynamics of recovery at presence of immunity in
the quarantine mode for fixed parameters $\beta=0.94$ and $A_{\max}=0.24$ and
different $A_{\min}=0.17, 0.1, 0.05, 0.02, 0.01$. One can see that now the
oscillations shown in Fig.\ref{Pic_Fig6a} acquire shape of strongly damped
picks that results the number of infected (order parameter $A$) to rapidly decrease. This
allows to predicts that after the first high pick of infection (that has nearly fixed amplitude for all the cases) may occur a second pick but with lesser amplitude and then the complete recover may become.%
   
\begin{figure}[ptb]%
\centering
\includegraphics[
width=0.5\textwidth
]%
{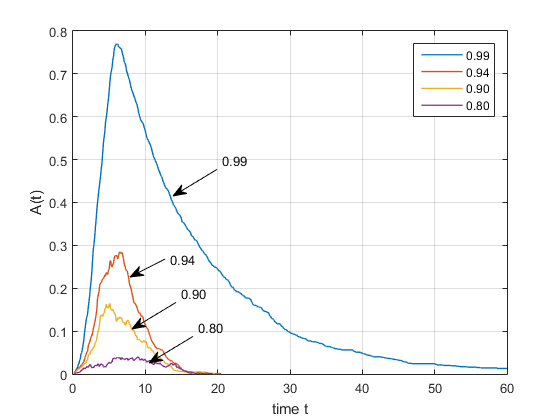}%
\caption{(Color on line.) The dynamics of infections (order parameter $A$) for different
values of the risk factor $\beta=0.99,0.94,0.9,0.8$ for situation when only the effective
personal immunity $M>0$ presents (without the quarantine), see Eq.(\ref{attrib}). This 
simulation shows that in such case the complete recover can occur for a lesser time comparing to Fig.\ref{Pic_Fig6a}.}%
\label{Pic_Fig7a}%
\end{figure}

Now we compare the effects of quarantine and immunity factors for
recovery. Fig.\ref{Pic_Fig7a} shows the dynamics of infections (order parameter $A$) for different
values of the risk factor $\beta=0.99, 0.94, 0.9, 0.8$ at situation without the quarantine 
when only the personal immunity $M>0$ presents (see Eq.(\ref{attrib})). This 
simulation shows that in such case the complete recover can occur even for a lesser time comparing to Fig.\ref{Pic_Fig6a}.

\section{Discussion and Conclusion}
We studied the dynamics of the infection spread at various values of the risk factors $\beta$ (control parameter) using the dynamic Monte Carlo method (DMC). In our model, it is accepted that the infection is transmitted through the contacts of randomly moving individuals. We show that the behavior of recovered individuals critically dependents on
the value $\beta$. For sub-critical values ​$\beta<\beta_{c}\sim 0.6$, the number of infected cases (the order parameters $A(t)$) asymptotically converges to zero, so that at moderate risk factor, the infection can quickly disappear.
\begin{figure}[ptb]%
	\centering
	\includegraphics[
	width=0.5\textwidth
	]%
	{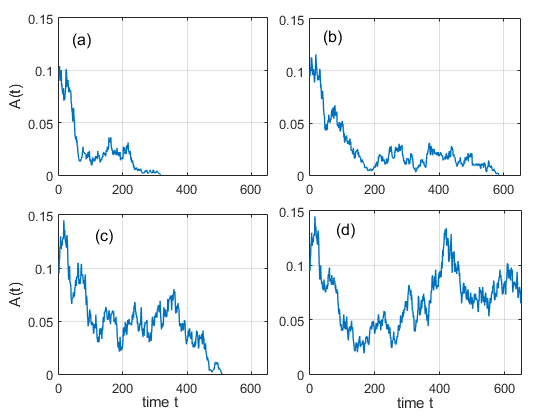}%
	\caption{(Color on line.)  The fraction of infections $A(t,\beta)$ as function of time $t$ at different  risk factors $\beta$ near the critical transition at $\beta_{c} \sim 0.6$ for $N=1000$ and initial number of infections $I_0=100$, with $\beta$: (a) 0.56,(b) 0.57,(c) 0.58,(d) 0.581. We observe that for $\beta \lesssim 0.58$ the number of infections rapidly reach zero. However for $\beta>0.8$ the process of recovering may be long.} 
	\label{Pic_Fig8a}%
\end{figure}
  However such a nontrivial behavior has to be confirmed by direct calculation.  Fig. \ref{Pic_Fig8a} shows the dynamics of infections fraction $A(t,\beta)$ with time for different risk factors $\beta$ near the critical transition $\beta \sim \beta_{c}=0.6$ for $N=1000$ and rather large the initial number of infections $I_0=100$. We observe that really for $\beta \lesssim 0.58$ the number of infections rapidly reach zero. We also analyzed the extended system, which currently is widely used to prevent the spread of the virus. In our approach such a system includes two additional parameters on/off the quarantine state. It was revealed that early exit from the quarantine leads to irregular oscillating dynamics (with positive Lyapunov exponent)  of the infection. However when the lower limit of the quarantine off is sufficiently small, the infection dynamics acquires a characteristic nonmonotonic shape with several damped peaks. The dynamics of the infection spread in case of individuals with immunity was studied too. Our comparison of the quarantine and the immunity factors on a recovery shows that in case of stable immunity a complete recovery occurs faster than in a quarantine mode.

\section{Acknowledgment}
This work was supported in part by CONACYT (M\'{e}xico) under the grant No. A1-S-9201.

\end{document}